\documentclass{emulateapj}

\newcommand{\Angstrom}{${\buildrel _{\circ} \over{\mathrm{A}}} \;$}

\shorttitle{UV Continuum Slopes}
\shortauthors{Kurczynski, P. et al.}

\received{2014 March 31}
\accepted{2014 July 25}
\slugcomment{Accepted for publication in the Astrophysical Journal Letters}

\usepackage{graphicx}

\DeclareGraphicsExtensions{.pdf,.png,.jpg,.eps}

\usepackage{amssymb, amsmath}

\usepackage[hyperindex,breaklinks]{hyperref}

\begin{document}


\title{The UV Continuum of $z>1$ Star-forming Galaxies in the Hubble Ultraviolet UltraDeep Field}


\author{Peter Kurczynski\altaffilmark{1}, Eric Gawiser\altaffilmark{1}, Marc Rafelski\altaffilmark{2}, Harry I. Teplitz\altaffilmark{3}, Viviana Acquaviva\altaffilmark{4}, Thomas M. Brown\altaffilmark{5}, Dan Coe\altaffilmark{5}, Duilia F. de Mello\altaffilmark{6,7},  Steven L. Finkelstein\altaffilmark{8}, Norman A. Grogin\altaffilmark{5}, Anton M. Koekemoer\altaffilmark{5}, Kyoung-soo Lee\altaffilmark{9}, Claudia Scarlata\altaffilmark{10}, Brian D. Siana\altaffilmark{11}}

\altaffiltext{1}{Department of Physics and Astronomy, Rutgers, The State University of New Jersey, Piscataway, NJ 08854, USA}
\altaffiltext{2}{NASA Postdoctoral Program Fellow, Goddard Space Flight Center, Code 665, Greenbelt, MD 20771, USA}
\altaffiltext{3}{Infrared Processing and Analysis Center, MS 100-22, Caltech, Pasadena, CA 91125, USA}
\altaffiltext{4}{New York City College of Technology, Brooklyn, NY 11201, USA} 
\altaffiltext{5}{Space Telescope Science Institute, 3700 San Martin Drive, Baltimore, MD 21218, USA}
\altaffiltext{6}{Laboratory for Observational Cosmology, Astrophysics Science Division, Code 665, Goddard Space Flight Center, Greenbelt, MD 20771, USA}
\altaffiltext{7}{Department of Physics, The Catholic University of America, Washington, DC 20064, USA}
\altaffiltext{8}{Department of Astronomy, The University of Texas at Austin, Austin, TX 78712, USA}
\altaffiltext{9}{Department of Physics, Purdue University, 525 Northwestern Avenue, West Lafayette, IN 47907, USA}
\altaffiltext{10}{Minnesota Institute for Astrophysics, School of Physics and Astronomy, University of Minnesota, Minneapolis, MN 55455, USA}
\altaffiltext{11}{Department of Physics and Astronomy, University of California, Riverside, CA 92521, USA}




%
%
\begin{abstract}
We estimate the UV continuum slope, $\beta$, for 923 galaxies in the range $1 < z < 8$ in the Hubble Ultradeep Field (HUDF).  These data include 460 galaxies at $1<z<2$ down to an absolute magnitude  $M_{\textnormal{UV}} = -14~(\sim 0.006~L^*_{z=1}; 0.02~L^*_{z=0}$), comparable to dwarf galaxies in the local universe.  We combine deep $HST$/UVIS photometry in $F225W$, $F275W$, $F336W$ wavebands (UVUDF) with recent data from $HST$/WFC3/IR (HUDF12).  Galaxies in the range $1<z<2$ are significantly bluer than local dwarf galaxies.  We find their mean (median) values $\left<\beta \right> = -1.382~(-1.830)\pm0.002$ (random) $\pm0.1$ (systematic).  We find comparable scatter in $\beta$ (standard deviation = 0.43) to local dwarf galaxies and 30\% larger scatter than $z>2$ galaxies.  We study the trends of $\beta$ with redshift and absolute magnitude for binned sub-samples and find a modest color-magnitude relation, $d\beta/dM = -0.11 \pm 0.01$ and no evolution in $d\beta/dM$ with redshift.  A modest increase in dust reddening with redshift and luminosity, $\Delta E(B-V) \sim 0.1$, and a comparable increase in the dispersion of dust reddening at $z<2$, appears likely to explain the observed trends.  At $z>2$, we find trends that are consistent with previous works; combining our data with the literature in the range $1<z<8$, we find a color evolution with redshift,  $d\beta/dz = -0.09\pm0.01$ for low luminosity (0.05 $L^*_{z=3}$), and  $d\beta/dz = -0.06\pm0.01$ for medium luminosity (0.25 $L^*_{z=3}$) galaxies.  
\end{abstract}

%
%
\section{Introduction}
\label{IntroductionSection}

The ultraviolet (UV) continuum provides an essential observational lever for understanding the formation and evolution of galaxies.  Understanding the trends of UV continuum slope with luminosity and redshift in galaxies bears upon their star formation rates, dust properties, stellar population ages and chemical enrichment.  Extending observations to the lowest luminosities is particularly important because these galaxies dominate the cosmic UV luminosity density and present significant challenges to models of galaxy evolution. 

Previously, UV continuum properties of high redshift galaxies have been studied with {\it Hubble Space Telescope (HST)} imaging in the redshift range $2 < z < 4$ down to absolute magnitudes $M_{\textnormal{UV}} \sim-18.5$ \citep{1999ApJ...521...64M,2004ApJ...600L.111P,2009ApJ...705..936B} and recently at $4 < z < 9$ down to absolute magnitudes $M_{\textnormal{UV}} \sim -17$ \citep{2011MNRAS.417..717W,2012ApJ...756..164F,2013arXiv1306.2950B,2013MNRAS.432.3520D,2013ApJ...772..136O}.  

Gravitational lensing has enabled studies of low luminosity ($M_{\textnormal{UV}} \sim -13$) high redshift galaxies \citep{2014ApJ...780..143A}; however, this technique has so far yielded small samples.  The {\it Galaxy Evolution Explorer} ($GALEX$; \citealt{2005ApJ...619L...1M}) has successfully studied the UV continuum of luminous $z<1$ galaxies (e.g \citealt{2011ApJ...726L...7O}), and ground-based studies have explored $z>1$ at higher luminosities (e.g. \citealt{2000ApJ...544..218A,2009ApJ...692..778R,2010ApJ...712.1070R,2011ApJ...733...99L,2012A&A...545A.141B}).

However, extending observations of the UV continuum to the lowest luminosity population of $z>1$ galaxies in large numbers has awaited the advent of deep field UV imaging with $HST$.  The Wide Field Camera (WFC3) Early Release Science (ERS; \citealt{2011ApJS..193...27W}) enabled detection of deep UV-selected galaxies in the range $1<z<3$  (\citealt{2010ApJ...720.1708H,2012ApJ...757...43H}, \citealt{2010ApJ...725L.150O}).  The Hubble Ultradeep Field (HUDF;  \citealt{2006AJ....132.1729B}) provides the deepest imaging currently available.  The most recent HUDF imaging in the near-IR (HUDF12; \citealt{2013ApJ...763L...7E}) and the UV (UVUDF; \citealt{2013AJ....146..159T}) enable the study of UV emission across the epoch of peak star formation, $1 < z < 4$, to fainter limits than ever before.

In this Letter, we utilize these data to probe the redshift range $1 < z < 8$ and measure the UV emission of galaxies with luminosities as low as $M_{\textnormal{UV}} \sim -14$ in the range $1<z<2$.  For comparison to the literature, we refer to luminosities in terms of $L^*_{z=3}$ reported in \citet{2009ApJ...692..778R}, $L^*_{z=1}$ reported in \citet{2004A&A...421...41G}, and $L^*_{z=0}$ reported in \citet{2011ApJS..192....6L}.  Magnitudes are in the AB system, and we adopt a cosmology with $\Omega_\Lambda$ = 0.7, $\Omega_0$ = 0.3, and $H_{0}$ = 70 km s$^{-1}$ Mpc$^{-1}$.

%
%
\section{Data and Sample Selection}
\label{SampleSection}
We utilize the 11 waveband $HST$ dataset for photometry, catalogs and redshifts.  Catalogs and photometry are made from images consisting of WFC3/UVIS mosaics in $F225W$, $F275W$, $F336W$ (UVUDF Epoch 3), ACS mosaics in $F435W$, $F606W$, $F775W$, $F850LP$, and WFC3/IR mosaics  in $F105W$, $F125W$, $F140W$, $F160W$. Objects are detected in a weighted sum of ACS and WFC3/IR images (8 images total) and catalogued based on aperture matched, PSF corrected photometry as described in \citet{2006AJ....132..926C} and Rafelski et al. (2014 in preparation).

The new UV photometry improves the accuracy of photometric redshift estimates, particularly at $z<3$.  Sampling the Lyman break reduces catastrophic errors and significantly improves redshift estimates \citep{2009ApJ...703.2033R}.  We use the Bayesian Photometric Redshift (BPZ) algorithm \citep{2000ApJ...536..571B} as described in \citet{2013ApJ...762...32C}.  There are 148 sources across the entire redshift range with ground-based spectroscopic confirmation, from which a photo-$z$ error is $\sigma_z = 1.8\%.$\footnote{$\sigma_z$ is the normalized median absolute deviation.  Given $dz \equiv |z_{\textnormal{spec}} - z_{\textnormal{phot}}|$, $\sigma_z \equiv 1.48~\times$ median $|(dz - \textrm{median}~dz)/(1+z_{\textnormal{spec}})|$ \citep{2008ApJ...686.1503B}.}  Nine nominal $10\sigma_z$ outliers (6.1\%) include seven sources that are near an image edge, have incomplete photometric coverage or segmentation problems.  The remaining two sources yield an outlier fraction of 1.35\%.  No outliers are included in our sample.  Previous photo-$z$ estimates in HUDF  \citep{2006AJ....132..926C}, have larger error (4.2\%) and nominal 10$\sigma_z$ outlier fraction (7.8\%) than the UVUDF.\footnote{We caution that \citet{2006AJ....132..926C} used different filters, especially in the IR, that may also affect a comparison.}  Details are presented in a forthcoming paper (Rafelski et al. 2014 in preparation).

Our sample selection criteria include:  
\begin{equation}
SN_{2330} > 5.0; z_{\textnormal{phot}} > 1.0; odds > 0.9; \chi^2_\nu < 2.0
\label{EQUATION:SampleSelectionCriteria}
\end{equation}

$SN_{2330} $ refers to signal-to-noise ratio at restframe 2330 \Angstrom, which is at the center of the wavelength range used for $\beta$ estimation (see below).  The odds requirement is a quality criterion imposed on the integral of the BPZ posterior probability distribution.  The reduced chi-squared, $\chi^2_\nu$, threshold rejects unreasonable fits.  We also reject fits where the estimated $\beta$ is at either end of the allowed range ($-10 < \beta < 10$), and by visual inspection (28 sources are rejected as unphysical SEDs or bad SED fits, 9 are likely photoz errors, 9 have segmentation problems, 6 are near an image edge or bright source and 1 is a known AGN).  The final sample consists of 923 galaxies, including 12 with spectroscopic confirmation (remaining spectroscopic sources do not meet the $\beta$ fit quality criteria or are at low redshift), that have apparent magnitudes $F435W > 23.6$ and span the redshift range $1.0 < z < 7.63$.

%
%
\section{Methods}
\label{MethodsSection}

The restframe UV continuum is characterized by a power law according to $f_\lambda \propto  \lambda^\beta;$ $\beta = -2$ corresponds to a flat spectrum in $f_\nu$.  Local starburst galaxies have $\left<\beta \right> \sim -1.3$;  as an extremely blue example, NGC 4861 has $\beta = -2.4$ \citep{1999ApJ...521...64M}.  The lowest expected value, $\beta=-3$, corresponds to the spectrum of individual O-type stars \citep{1995ApJS...96....9L}.  

The wavelength range [1260,2600]~\Angstrom is widely adopted for $\beta$ estimation \citep{1994ApJ...429..582C}.  We use an expanded range,  [1260,3400]~\Angstrom restframe, and we find that fits using this range have fewer $\beta$ outliers, smaller 68\% confidence errors, and a broader, more plausible distribution of $\chi^2$ values in the range $1<z<3$.  Comparison of $\beta$ estimates for galaxies in our sample that also appear in \citet{2012ApJ...756..164F}, using identical photometry, shows a marginal systematic difference, $\left< \Delta \beta \right> = 0.09 $ and scatter, $\sigma$ = 0.19, between the two methods.  

To avoid contamination from Ly$-\alpha$ emission or the Ly$-\alpha$ decrement, we adjust the blue end of the fitting range slightly to exclude photometry for galaxies where the Ly$-\alpha$ line would fall within the full width tenth maximum of the system throughput for the bluest filter in the fitting wavelength range.

The spectral energy distributions (SEDs) of galaxies in the entire catalog are fit using $\chi^2$ minimization in order to determine their UV spectral indices, $\beta$.  We use power-law templates that span the range $\beta \sim [-10,10]$ with stepsize = 0.01.  For each template, the spectrum is multiplied by $HST$ system throughputs from $F225W$ through $F160W$ wavebands to yield predicted photometry modulo a normalization factor.  Normalizations are determined analytically for each fit.  We require at least three points in the fits for $\beta$ estimation, and we estimate $M_{\textnormal{UV}}$ at 2330 \Angstrom,  in the center of the wavelength range used for $\beta$ estimation to avoid bias in the $M_{\textnormal{UV}}$ estimates.


For each galaxy in the sample, absolute magnitude at restframe 2330 \Angstrom and 1500 \Angstrom (for comparison to literature) is computed by integrating the best-fit model spectrum over a centered, rectangular 100 \Angstrom bandpass.  Average $M_{2330}$ values are found to be 0.1 magnitude more luminous than $M_{1500}$ values (standard deviation = 0.2 magnitude).

We determine errors to the $\beta$ and  $M_{2330}$ estimates from Monte Carlo simulation.  Simulated sources span the range of redshifts, $\beta$ and $M_{2330}$ values encountered in the data.  Photometry  is simulated in the observed wavebands by multiplying model spectral flux densities by system throughputs for each filter.  Errors are drawn at random from magnitude-matched sources in the actual data.  Scatter and bias of the $\beta$ estimates are determined from simulations with 10$^{3}$ realizations.  Errors in $\beta$ are found to be a few percent for the brightest, and most red sources and increase toward 20\% for the faintest and bluest sources in our sample.  We find negligible bias.  
%
%
%
\section{Results}
\label{ResultsSection}

The deep HUDF data reveal galaxies down to absolute magnitude $M_{\textnormal{UV}} = -14$.  The lowest luminosity galaxies are found in the redshift range $1<z\leq2$, where they sample the faint end of the UV luminosity function at $z=1$ ($0.006L^*_{z=1}$; \citealt{2004A&A...421...41G}), and are comparable to the bulk of the dwarf galaxy population in the local universe \citep{2011ApJS..192....6L}.  

In addition to improving the redshift accuracy as discussed above, the new, deep UV photometry, particularly for galaxies in the range $1<z<3$, is important in several respects.  First, the UV photometry increases sample size; without the UV photometry, the number of sources with $M_{\textnormal{UV}} > -16$ meeting our photo-$z$ odds criterion would be reduced by a factor of $\sim 2$.   Second, the $F275W$ and $F336W$ bands are essential for estimating $\beta$ with high fidelity (more than two data points in the fitting wavelength range) at $z\approx1$.

Across the entire redshift range, we find UV spectral indices that are on the whole bluer than those found in the local universe.  At $1<z<2$, we find mean (median) value $\left<\beta\right> = -1.382~(-1.800) \pm 0.002$ (random error), and standard deviation = 0.43, see Table \ref{TABLE:ResultsSummary}.  In comparison, a $GALEX$ sample of local dwarf galaxies has average (median) values of $\beta = -1.15~(-1.29)$\footnote{$\beta$ values are computed from reported $GALEX$ FUV-NUV colors using the method of \citet{2006ApJ...637..242C}.} with standard deviation 0.48 \citep{2010AJ....139..447H}.

\begin{figure}[h]
\begin{center}
\includegraphics[scale=0.4]{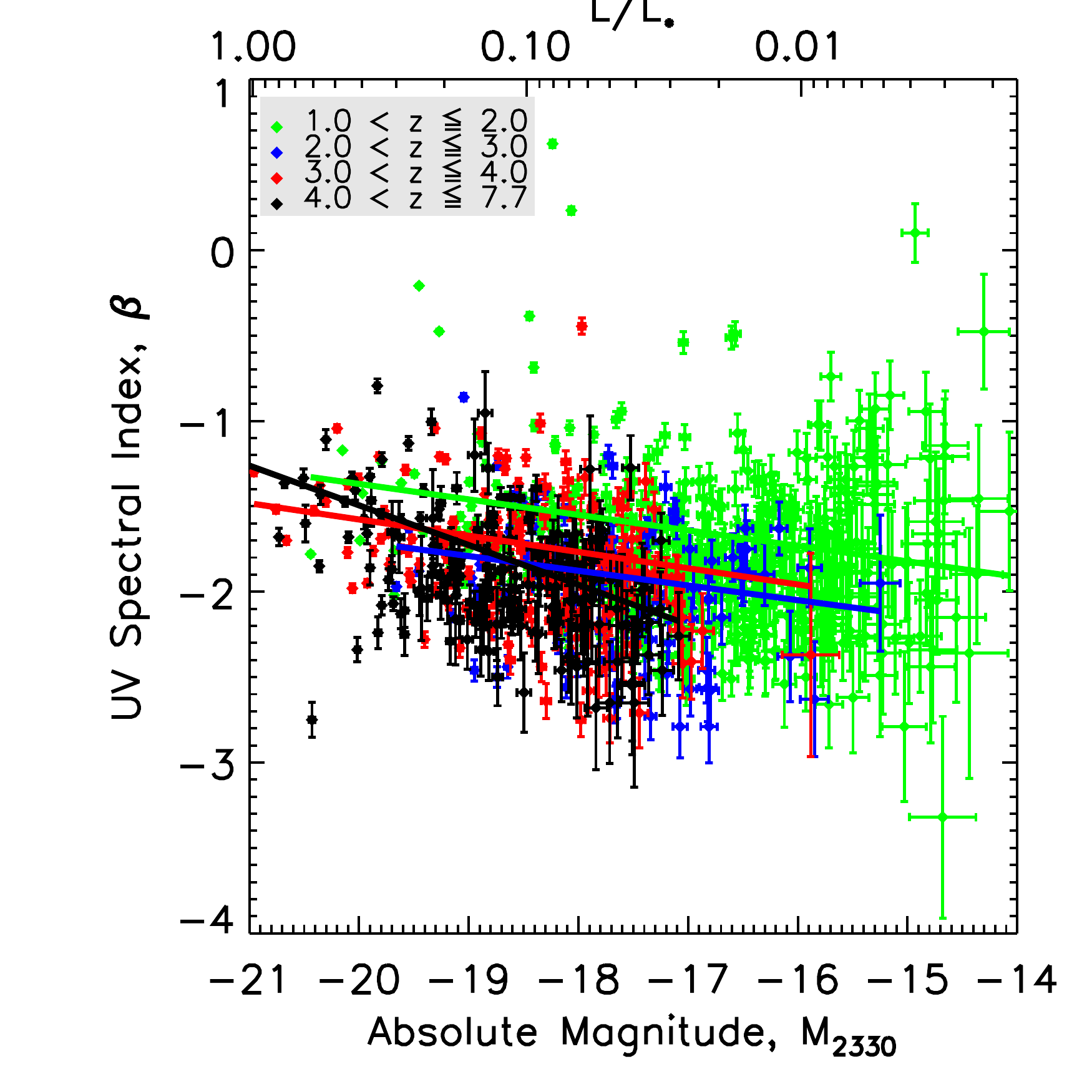}
\end{center}
\caption{UV spectral index, $\beta$, vs. absolute magnitude for galaxies in the UVUDF.  Errors are determined from simulation.  The top horizontal axis indicates the luminosity ratio, $L/L_*$, based on the luminosity function at $z\sim3$ in \citet{2009ApJ...692..778R}.  Galaxies are binned in redshift ranges as indicated in the legend. Linear fits to redshift-binned subsamples are shown as solid lines of corresponding colors.  Outliers that include three galaxies with $\beta >0$ have no obvious faults, such as bad segmentation or appearance near an edge in the images.}  
\label{FIGURE:BetaVsMagnitude}
\end{figure}

A modest color magnitude ($\beta-M$) trend, whereby more luminous galaxies tend to be redder (larger $\beta$) on average has been reported for galaxies spanning a range of redshifts (\citealt{2004ApJ...600L.111P}, and references therein, \citealt{2004ApJ...611..660O}, \citealt{2007ApJ...665..944L}, \citealt{2011MNRAS.417..717W}, \citealt{2009ApJ...705..936B,2012ApJ...754...83B,2013arXiv1306.2950B}).  Our redshift binned sub-samples show correlations (Spearman coefficient, $\rho$) in the range $-0.37 < \rho < -0.16$ with varying significance ($\sim10^{-2}$), see Figure \ref{FIGURE:BetaVsMagnitude} and Table \ref{TABLE:ResultsSummary}.

We find a larger scatter in $\beta$ at $z<2$ than at higher redshifts.   For galaxies with $M_{\textnormal{UV}} > -16$ incompleteness and large errors in $\beta$ may bias a comparison with $z>2$; however, even $M_{\textnormal{UV}}  < -16$ galaxies in the range $1<z<2$ have $\approx$30\% greater dispersion in $\beta$ compared to $z>2$ galaxies.  

We estimate the slopes, $d\beta/dM$, for redshift binned sub-samples by fitting to a linear model, $\beta = a + d\beta/dM \times M$. The estimates are formally not well constrained ($\chi^2$ probability $< 0.1$; \citealt{NumRecipes,Bevington}) due to intrinsic scatter.  Nevertheless, we adopt the best fit values in further analyses, and determine errors from bootstrap simulation.  

\begin{figure}[h]
\begin{center}
\includegraphics[scale=0.40]{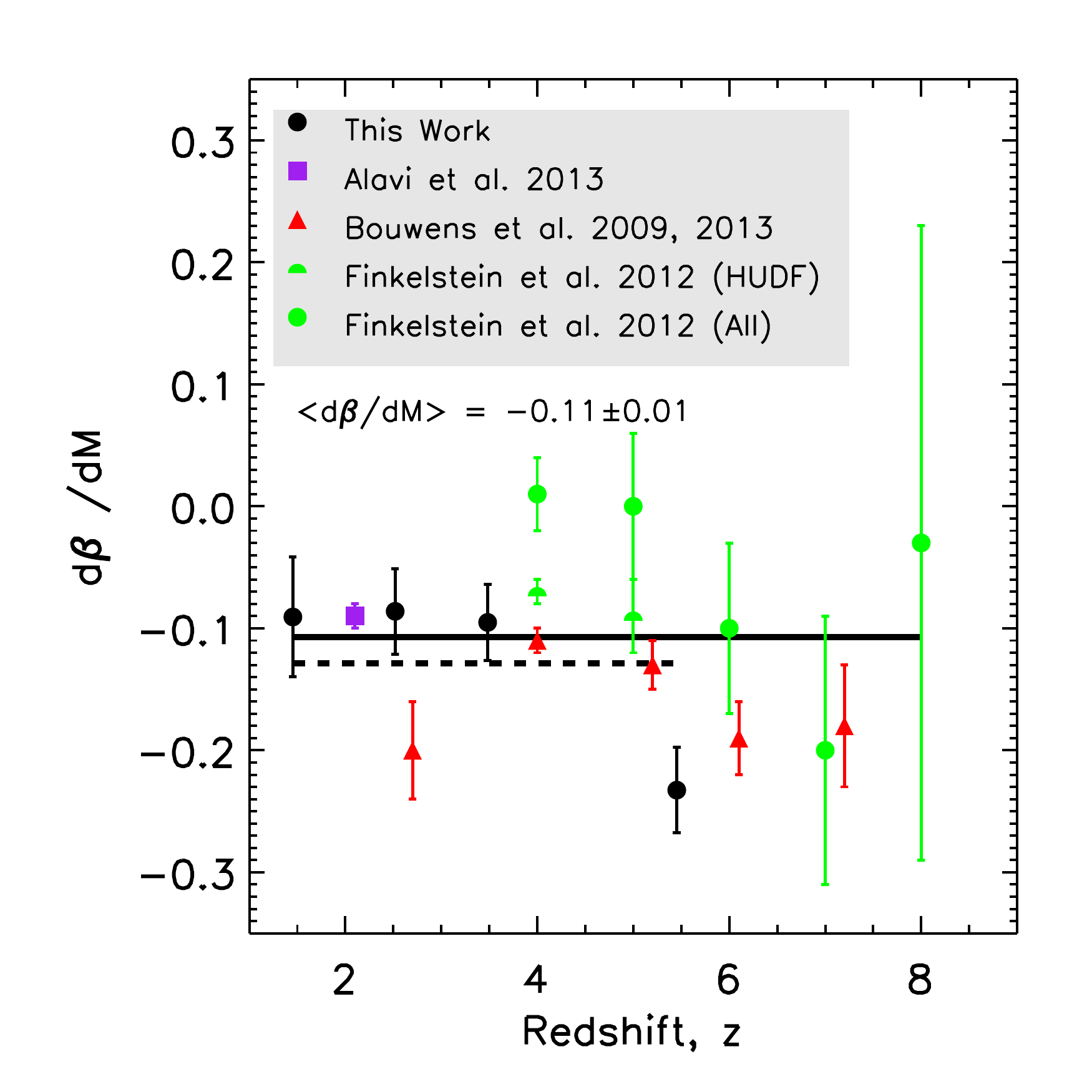}

\end{center}
\caption{Derivative of the UV spectral index with magnitude, $d\beta / dM$, vs. redshift for UVUDF galaxies compared to literature.  Filled, black circles refer to the present work, with galaxies binned in redshift, as in Figure \ref{FIGURE:BetaVsMagnitude}. Random errors are determined from bootstrap resampling simulation.  Sub-sample mean redshift is used for placement on the horizontal axis.  Points from the literature are indicated in the legend.  Overlapping points have been offset in redshift slightly for clarity.  The weighted average $d\beta/dM$ for the present data (combined literature) is indicated as a dashed (solid) black line.}
\label{FIGURE:dbdmVsMUVLiterature}
\end{figure}

We consider the color-magnitude relation across the redshift range $1.5 < z < 8$.  Figure \ref{FIGURE:dbdmVsMUVLiterature} shows our results combined with previously reported values of $d\beta/dM$\citep{2009ApJ...705..936B,2012ApJ...756..164F,2014ApJ...780..143A,2013arXiv1306.2950B}.  These combined results have an inverse variance weighted average value $\left<d\beta/dM\right> = -0.11 \pm 0.01$ (standard error of the mean), standard deviation = 0.08.  The combined literature do not reveal evolution in $d\beta/dM$ with redshift, even though somewhat different ranges of M$_{UV}$ have been used to determine $d\beta/dM$ at each redshift. 

\begin{figure}[h]
\begin{center}

\includegraphics[scale=0.40]{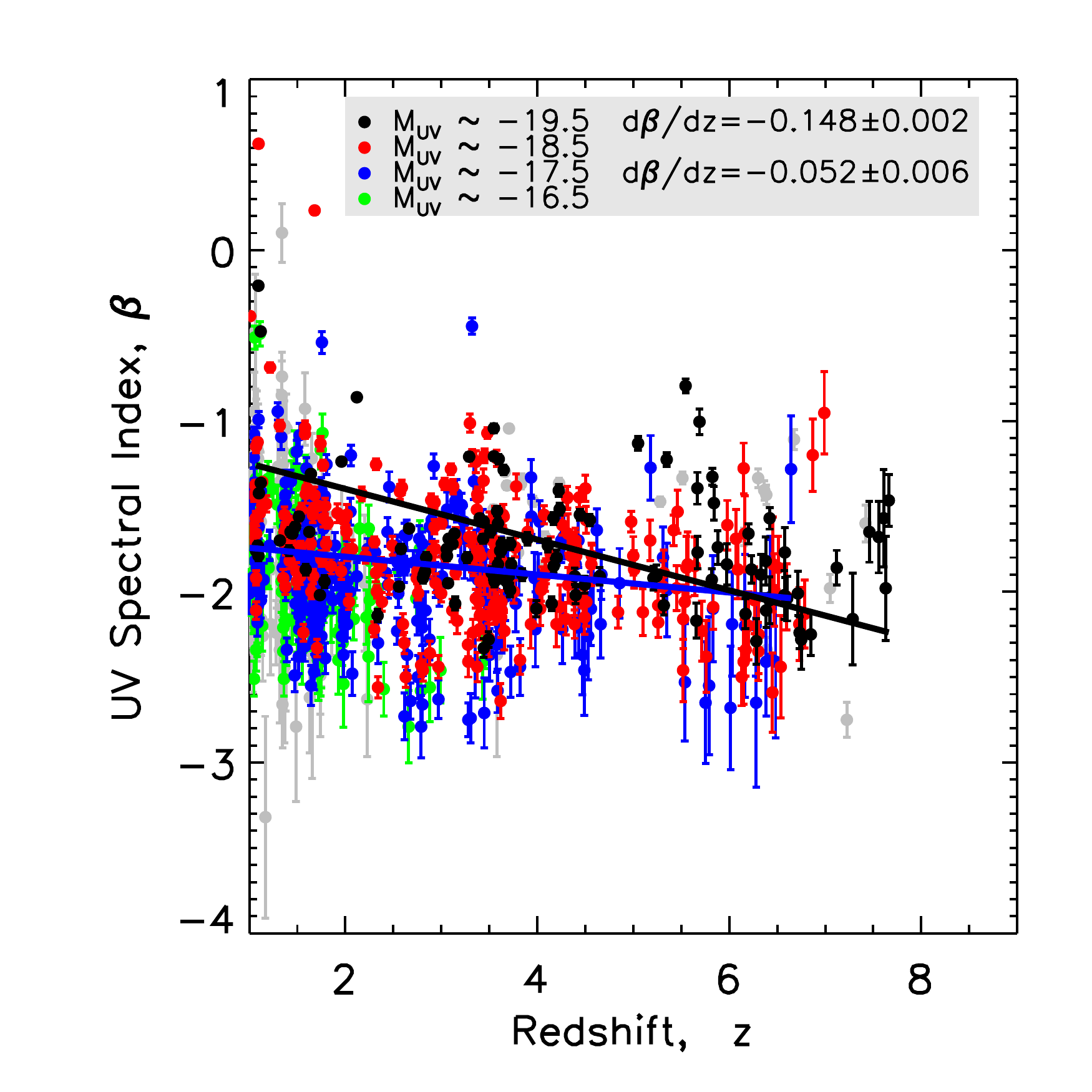}

\end{center}
\caption{UV spectral index, $\beta$, vs redshift for galaxies in the UVUDF.  Luminosity binned sub-samples ($-20 < M_{\textnormal{UV}} < -19$; black, $-19 < M_{\textnormal{UV}} < -18$; red, $-18 < M_{\textnormal{UV}} < -17$; blue, $M_{\textnormal{UV}} > -17$; green) sub-samples are shown superimposed on the remaining sample (gray) that include galaxies at either end of the luminosity distribution.  Fits to representative sub-samples are shown as lines with slopes, $d\beta/dz$, indicated in the legend.}

\label{FIGURE:BetaVsRedshift}
\end{figure}

Galaxies tend to be bluer at higher redshift, and we investigate the trend of $\beta$ versus $z$ in Figure \ref{FIGURE:BetaVsRedshift}.  We estimate the slopes, $d\beta/dz$, for magnitude binned sub-samples by fitting to a linear model, $\beta = a + d\beta/dz \times z$.  We report the slope values with caution since they are also not formally well constrained ($\chi^2$ probability $< 0.1$).  

For comparison with the literature, we select bins of low luminosity ( $M_{\textnormal{UV}}\sim -17.5$; $-18 < M_{\textnormal{UV}} < -17$) and medium luminosity ($M_{\textnormal{UV}}\sim-19.5$; $-20<M_{\textnormal{UV}}<-19$) galaxies that are also binned in redshift ($\Delta z = 1.0$, $\approx10$ galaxies/bin, medium luminosity; $\Delta z = 0.5$, $1<z<4$ and $\Delta z = 1.0$ $4<z<6$ $\approx30$ galaxies/bin, low luminosity), see Figure \ref{FIGURE:BetaRedshiftLiterature}.  We find a significant correlation of $\beta$ with redshift, see Table \ref{TABLE:ResultsSummary}, and significant $d\beta/dz=-0.09~(0.06) \pm 0.01$ for $M_{\textnormal{UV}} \sim -17.5~( -19.5)$ subsamples respectively.  

To investigate systematics, we repeated our analysis with variants of our sample selection criteria.  We varied each criterion and repeated the analysis.  We changed the wavelength range from [1260,3400] \Angstrom to [1260,2600] \Angstrom, we relaxed the BPZ odds threshold from odds $> 0.9$ to odds $> 0.5$, and we relaxed the reduced $\chi^2$ threshold from $\chi^2_\nu < 2.0$ to $\chi^2_\nu < 5.0$.  The mean and median $\beta$ were found to decrease by 10\% (i.e. blue ward) when relaxing the BPZ odds requirement; other variations had negligible effect on the mean and the median $\beta$.  Changing the wavelength range increased $d\beta/dM$ by 0.1, but relaxing the BPZ odds requirement or the reduced $\chi^2$ threshold caused negligible changes in slope.  Accordingly, we estimate systematic uncertainties of 0.1 in $d\beta/dM$ and 0.1 in $\left< \beta \right>$.  

The evolution of $\beta$ is affected by dust reddening, age of the stellar populations and metallicity.  To interpret the trends in $\beta$, we compare the observations to stellar population synthesis models.  Inspecting \citet{2003MNRAS.344.1000B} models with two of these three parameters fixed to the values $E(B-V) = 0.25$, age = 30 Myr, metallicity, $0.2Z_\odot$ and the third parameter varied, we find variations of $\beta$ with these parameters of $d\beta/d~E(B-V) = 4.2$, $d\beta/d~\textrm{age} = 0.71$ Gyr$^{-1}$, $d\beta/d~log Z = 0.26$ (metallicity in units of $Z_\odot$). 

The modest evolution in $\beta$ seen from $z=3$ to $z=1$ ($\Delta \beta = 0.30$), and from $M_{\textnormal{UV}} \sim-14$ to $M_{\textnormal{UV}} \sim-20$ at $z\sim2$ ($\Delta \beta = 0.36$) can be explained by a small increase of $\sim 0.1$  in $E(B-V)$,\footnote{We interpret changes in $E(B-V)$ as solely due to dust; Alavi et al. (2014) find no dependence of $E(B-V)$ on age, although a modest dependence is found on metallicity for low luminosity, $z\sim2$ galaxies.}  a $\sim$0.4 Gyr increase in average stellar population age, or a factor of $\sim$14 increase in metallicity.  Such large variation in typical metallicity seems unlikely in 3 Gyr of cosmic time or across a factor of 100 in luminosity.  An increase in average age is less implausible.  Although neither trend can be ruled out, and all three properties are likely evolving in unison, we hypothesize that the dominant source of evolution in $\beta$ is a modest increase in dust reddening as a function of time and rest-UV luminosity.  Similarly, the observed scatter in $\beta$ at $1<z<2$ ($\sigma = 0.43$) can be explained by a dispersion of 0.10 in $E(B-V)$,  a dispersion in age of 2~Gyr, or a factor of 10 dispersion in metallicity over this redshift range, which follows the peak in cosmic star formation at $z\sim2$.  

%
\section{Conclusion}
\label{ConclusionSection}
Deep UVUDF photometry enables the study of galaxies at $1<z<2$ down to an absolute magnitude, $M_{\textnormal{UV}} = -14~(\sim 0.006~L^*_{z=1}; 0.02~L^*_{z=0})$, comparable to dwarf galaxies in the local universe.   These galaxies have mean $\beta$ values comparable to local starbursts, rather than local dwarf galaxies.  

We find a modest color-magnitude relation, qualitatively similar to previous literature, whereby less luminous galaxies tend to be bluer, $d\beta/dM = -0.11 \pm 0.01$ (random) $\pm 0.1$ (systematic).  Combining these data with literature at higher redshift, we do not find evidence for evolution in this color-magnitude relation in the range $1 < z < 8$.  

We find significant color evolution with redshift; galaxies with $M_{\textnormal{UV}} \sim -19.5$ ($0.25 L^*_{z=3}$) have $d\beta/dz = -0.06 \pm 0.01$ (significance $\sim10^{-5}$) in the range $1 < z < 8$, from combining current data with the literature.  Lower luminosity galaxies ($M_{\textnormal{UV}} \sim -17.5$) show a similar trend, $d\beta/dz = -0.09\pm0.01$, in the combined literature.  

The trends of $\beta$ with redshift and magnitude in the range $1<z<3$ can be explained by increased dust attenuation, $\Delta E(B-V) \sim 0.1$, corresponding to $\approx$ 1 magnitude assuming a \citet{2000ApJ...533..682C} attenuation law.  Alternatively, a ~0.4 Gyr increase in average stellar population age, or a factor of $\sim$14 increase in metallicity can explain the observations.  The observed scatter in $\beta$ can be similarly explained by comparable dispersions in dust, age or metallicity.

\begin{figure}[h]
\begin{center}

\includegraphics[scale=0.35]{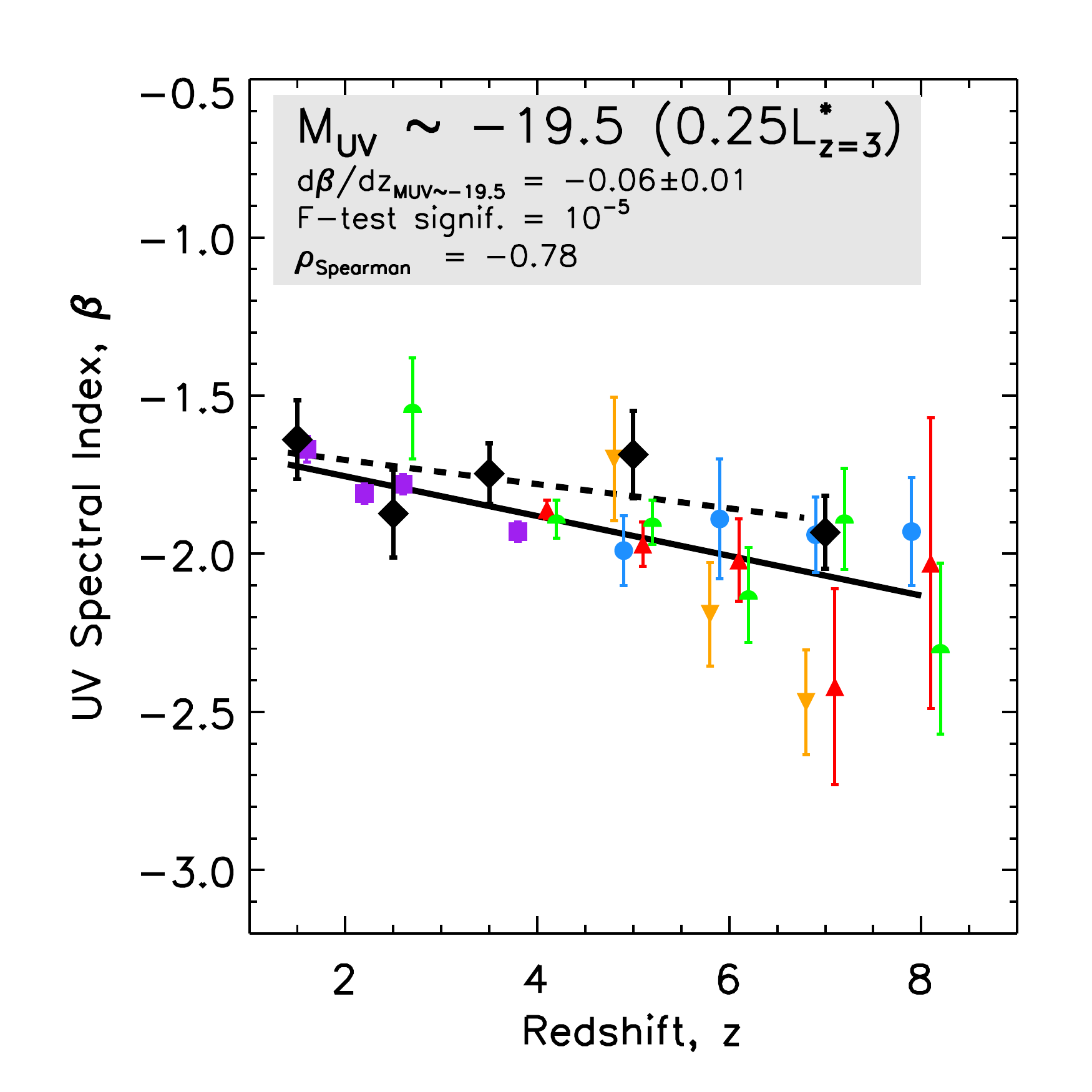}
\includegraphics[scale=0.35]{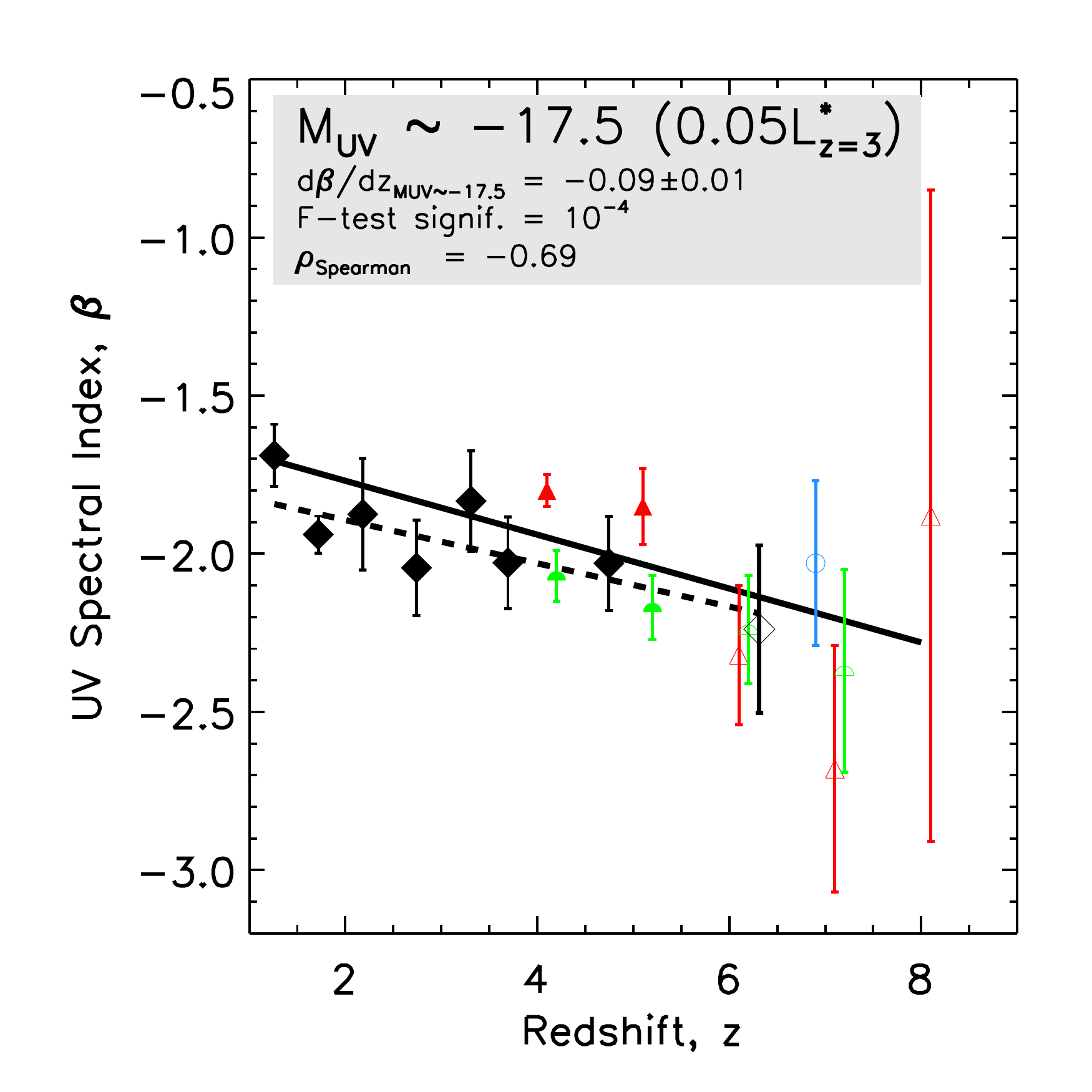}

\end{center}
\caption{UV spectral index, $\beta$, vs. redshift for medium luminosity (left panel; M$_{UV} \sim -19.5$; $0.25 L_{z=3}^*$) and low luminosity (right panel; M$_{UV} \sim -17.5$; $0.05 L_{z=3}^*$) galaxies in the UVUDF, combined with the literature.  Present data are shown as black diamonds.  Literature data are from Bouwens et al. 2009, 2013 (green half-circles), Finkelstein et al. 2012 (red upward pointing triangles), Hathi et al. 2013 (purple squares), Dunlop et al. 2012, 2013 (blue circles) and Wilkins et al. 2011 (orange downward pointing triangles).  Open symbols in the right panel denote samples that are expected to be less than  50\% complete.  Linear fits to the present data (combined literature) are illustrated as dashed (solid) lines.  Legend statistics correspond to the combined literature.}
\label{FIGURE:BetaRedshiftLiterature}
\end{figure}

Dust also leads to significant FIR-sub-millimeter luminosity as dust absorbed starlight is thermally re-radiated.  The trends in Figure \ref{FIGURE:BetaRedshiftLiterature} place upper limits on the accumulation of dust  spanning the epoch of peak cosmic star formation, $1<z<4$.   Using observed $\beta$ values and the empirical IRX-$\beta$ relationship of \citet{1999ApJ...521...64M} to estimate the dust-induced IR luminosities, we find the comoving dust luminosity density of typical (M$_{UV} \sim -19.5$) galaxies to grow by a factor of $\sim10$ from $z=4$ to $z=1.$  Aging stellar populations and changes in metallicity would also increase $\beta$; therefore this growth factor must be interpreted as a rough upper limit. 

In general, we do not find $\beta$ values as low as -3, even in the lowest luminosity galaxies.  Maximally blue values ($\beta \approx -3$) could have been interpreted as due to the infall of pristine gas and exclusively young stars, since dust, metallicity and aging all {\it redden} spectra \citep{2003ApJ...591..878D}.  The absence of such blue values of $\beta$ may support scenarios of galaxy evolution that invoke accretion of metal-rich gas, as opposed to pristine gas, although other scenarios involving dust accumulation can also be supported.

The estimation of stellar masses, metallicities, star formation rates and histories of these low luminosity galaxies, enabled by the wealth of data in the HUDF, will contribute to a more complete synthesis of the formation and evolution of galaxies.

\vspace{0.5in}

The authors wish to thank the anonymous referee for comments that greatly improved this manuscript.  Support for HST Program GO-12534 was provided by NASA through grants from the Space Telescope Science Institute, which is operated by the Association of Universities for Research in Astronomy, Inc., under NASA contract NAS5-2655.  This material is based upon work supported by the National Science Foundation under grant no. 1055919.
\bibliographystyle{apj}                       

%
%
%
%

\bibliography{uvudf_beta_estimation_bibliography.bib}

\begin{thebibliography}{42}
\expandafter\ifx\csname natexlab\endcsname\relax\def\natexlab#1{#1}\fi

\bibitem[{{Adelberger} \& {Steidel}(2000)}]{2000ApJ...544..218A}
{Adelberger}, K.~L. \& {Steidel}, C.~C. 2000, \apj, 544, 218

\bibitem[{{Alavi} {et~al.}(2014){Alavi}, {Siana}, {Richard},
  {et~al.}}]{2014ApJ...780..143A}
{Alavi}, A., {Siana}, B., {Richard}, J., {et~al.} 2014, \apj, 780, 143

\bibitem[{{Beckwith} {et~al.}(2006){Beckwith}, {Stiavelli}, {Koekemoer},
  {et~al.}}]{2006AJ....132.1729B}
{Beckwith}, S.~V.~W., {Stiavelli}, M., {Koekemoer}, A.~M., {et~al.} 2006, \aj,
  132, 1729

\bibitem[{{Ben{\'{\i}}tez}(2000)}]{2000ApJ...536..571B}
{Ben{\'{\i}}tez}, N. 2000, \apj, 536, 571

\bibitem[{{Bevington} \& {Robinson}(1992)}]{Bevington}
{Bevington}, P.~R. \& {Robinson}, D.~K. 1992, Data Reduction and Error Analysis
  For the Physical Sciences (New York: McGraw-Hill Inc.)

\bibitem[{{Bouwens} {et~al.}(2009){Bouwens}, {Illingworth}, {Franx},
  {et~al.}}]{2009ApJ...705..936B}
{Bouwens}, R.~J., {Illingworth}, G.~D., {Franx}, M., {et~al.} 2009, \apj, 705,
  936

\bibitem[{{Bouwens} {et~al.}(2012){Bouwens}, {Illingworth}, {Oesch},
  {et~al.}}]{2012ApJ...754...83B}
{Bouwens}, R.~J., {Illingworth}, G.~D., {Oesch}, P.~A., {et~al.} 2012, \apj,
  754, 83

\bibitem[{{Bouwens} {et~al.}(2013){Bouwens}, {Illingworth}, {Oesch},
  {et~al.}}]{2013arXiv1306.2950B}
---. 2013, ArXiv e-prints 1306.2950

\bibitem[{{Brammer} {et~al.}(2008){Brammer}, {van Dokkum}, \&
  {Coppi}}]{2008ApJ...686.1503B}
{Brammer}, G.~B., {van Dokkum}, P.~G., \& {Coppi}, P. 2008, \apj, 686, 1503

\bibitem[{{Bruzual} \& {Charlot}(2003)}]{2003MNRAS.344.1000B}
{Bruzual}, G. \& {Charlot}, S. 2003, \mnras, 344, 1000

\bibitem[{{Buat} {et~al.}(2012){Buat}, {Noll}, {Burgarella},
  {et~al.}}]{2012A&A...545A.141B}
{Buat}, V., {Noll}, S., {Burgarella}, D., {et~al.} 2012, \aap, 545, A141

\bibitem[{{Calzetti} {et~al.}(2000){Calzetti}, {Armus}, {Bohlin},
  {et~al.}}]{2000ApJ...533..682C}
{Calzetti}, D., {Armus}, L., {Bohlin}, R.~C., {et~al.} 2000, \apj, 533, 682

\bibitem[{{Calzetti} {et~al.}(1994){Calzetti}, {Kinney}, \&
  {Storchi-Bergmann}}]{1994ApJ...429..582C}
{Calzetti}, D., {Kinney}, A.~L., \& {Storchi-Bergmann}, T. 1994, \apj, 429, 582

\bibitem[{{Coe} {et~al.}(2006){Coe}, {Ben{\'{\i}}tez}, {S{\'a}nchez},
  {et~al.}}]{2006AJ....132..926C}
{Coe}, D., {Ben{\'{\i}}tez}, N., {S{\'a}nchez}, S.~F., {et~al.} 2006, \aj, 132,
  926

\bibitem[{{Coe} {et~al.}(2013){Coe}, {Zitrin}, {Carrasco},
  {et~al.}}]{2013ApJ...762...32C}
{Coe}, D., {Zitrin}, A., {Carrasco}, M., {et~al.} 2013, \apj, 762, 32

\bibitem[{{Cortese} {et~al.}(2006){Cortese}, {Boselli}, {Buat},
  {et~al.}}]{2006ApJ...637..242C}
{Cortese}, L., {Boselli}, A., {Buat}, V., {et~al.} 2006, \apj, 637, 242

\bibitem[{{Dorman} {et~al.}(2003){Dorman}, {O'Connell}, \&
  {Rood}}]{2003ApJ...591..878D}
{Dorman}, B., {O'Connell}, R.~W., \& {Rood}, R.~T. 2003, \apj, 591, 878

\bibitem[{{Dunlop} {et~al.}(2013){Dunlop}, {Rogers}, {McLure},
  {et~al.}}]{2013MNRAS.432.3520D}
{Dunlop}, J.~S., {Rogers}, A.~B., {McLure}, R.~J., {et~al.} 2013, \mnras, 432,
  3520

\bibitem[{{Ellis} {et~al.}(2013){Ellis}, {McLure}, {Dunlop},
  {et~al.}}]{2013ApJ...763L...7E}
{Ellis}, R.~S., {McLure}, R.~J., {Dunlop}, J.~S., {et~al.} 2013, \apjl, 763, L7

\bibitem[{{Finkelstein} {et~al.}(2012){Finkelstein}, {Papovich}, {Salmon},
  {et~al.}}]{2012ApJ...756..164F}
{Finkelstein}, S.~L., {Papovich}, C., {Salmon}, B., {et~al.} 2012, \apj, 756,
  164

\bibitem[{{Gabasch} {et~al.}(2004){Gabasch}, {Bender}, {Seitz},
  {et~al.}}]{2004A&A...421...41G}
{Gabasch}, A., {Bender}, R., {Seitz}, S., {et~al.} 2004, \aap, 421, 41

\bibitem[{{Hathi} {et~al.}(2012){Hathi}, {Mobasher}, {Capak}, {Wang}, \&
  {Ferguson}}]{2012ApJ...757...43H}
{Hathi}, N.~P., {Mobasher}, B., {Capak}, P., {Wang}, W.-H., \& {Ferguson},
  H.~C. 2012, \apj, 757, 43

\bibitem[{{Hathi} {et~al.}(2010){Hathi}, {Ryan}, {Cohen},
  {et~al.}}]{2010ApJ...720.1708H}
{Hathi}, N.~P., {Ryan}, Jr., R.~E., {Cohen}, S.~H., {et~al.} 2010, \apj, 720,
  1708

\bibitem[{{Hunter} {et~al.}(2010){Hunter}, {Elmegreen}, \&
  {Ludka}}]{2010AJ....139..447H}
{Hunter}, D.~A., {Elmegreen}, B.~G., \& {Ludka}, B.~C. 2010, \aj, 139, 447

\bibitem[{{Labb{\'e}} {et~al.}(2007){Labb{\'e}}, {Franx}, {Rudnick},
  {et~al.}}]{2007ApJ...665..944L}
{Labb{\'e}}, I., {Franx}, M., {Rudnick}, G., {et~al.} 2007, \apj, 665, 944

\bibitem[{{Lee} {et~al.}(2011{\natexlab{a}}){Lee}, {Gil de Paz}, {Kennicutt},
  {et~al.}}]{2011ApJS..192....6L}
{Lee}, J.~C., {Gil de Paz}, A., {Kennicutt}, Jr., R.~C., {et~al.}
  2011{\natexlab{a}}, \apjs, 192, 6

\bibitem[{{Lee} {et~al.}(2011{\natexlab{b}}){Lee}, {Dey}, {Reddy},
  {et~al.}}]{2011ApJ...733...99L}
{Lee}, K.-S., {Dey}, A., {Reddy}, N., {et~al.} 2011{\natexlab{b}}, \apj, 733,
  99

\bibitem[{{Leitherer} \& {Heckman}(1995)}]{1995ApJS...96....9L}
{Leitherer}, C. \& {Heckman}, T.~M. 1995, \apjs, 96, 9

\bibitem[{{Martin} {et~al.}(2005){Martin}, {Fanson}, {Schiminovich},
  {et~al.}}]{2005ApJ...619L...1M}
{Martin}, D.~C., {Fanson}, J., {Schiminovich}, D., {et~al.} 2005, \apjl, 619,
  L1

\bibitem[{{Meurer} {et~al.}(1999){Meurer}, {Heckman}, \&
  {Calzetti}}]{1999ApJ...521...64M}
{Meurer}, G.~R., {Heckman}, T.~M., \& {Calzetti}, D. 1999, \apj, 521, 64

\bibitem[{{Oesch} {et~al.}(2010){Oesch}, {Bouwens}, {Carollo},
  {et~al.}}]{2010ApJ...725L.150O}
{Oesch}, P.~A., {Bouwens}, R.~J., {Carollo}, C.~M., {et~al.} 2010, \apjl, 725,
  L150

\bibitem[{{Oesch} {et~al.}(2013){Oesch}, {Labb{\'e}}, {Bouwens},
  {et~al.}}]{2013ApJ...772..136O}
{Oesch}, P.~A., {Labb{\'e}}, I., {Bouwens}, R.~J., {et~al.} 2013, \apj, 772,
  136

\bibitem[{{Ouchi} {et~al.}(2004){Ouchi}, {Shimasaku}, {Okamura},
  {et~al.}}]{2004ApJ...611..660O}
{Ouchi}, M., {Shimasaku}, K., {Okamura}, S., {et~al.} 2004, \apj, 611, 660

\bibitem[{{Overzier} {et~al.}(2011){Overzier}, {Heckman}, {Wang},
  {et~al.}}]{2011ApJ...726L...7O}
{Overzier}, R.~A., {Heckman}, T.~M., {Wang}, J., {et~al.} 2011, \apjl, 726, L7

\bibitem[{{Papovich} {et~al.}(2004){Papovich}, {Dickinson}, {Ferguson},
  {et~al.}}]{2004ApJ...600L.111P}
{Papovich}, C., {Dickinson}, M., {Ferguson}, H.~C., {et~al.} 2004, \apjl, 600,
  L111

\bibitem[{{Press} {et~al.}(1990){Press}, {Flannery}, {Teukolsky}, \&
  {Vetterling}}]{NumRecipes}
{Press}, W.~H., {Flannery}, B.~P., {Teukolsky}, S.~A., \& {Vetterling}, W.~T.
  1990, Numerical Recipes in C (New York: Cambridge University Press)

\bibitem[{{Rafelski} {et~al.}(2009){Rafelski}, {Wolfe}, {Cooke},
  {et~al.}}]{2009ApJ...703.2033R}
{Rafelski}, M., {Wolfe}, A.~M., {Cooke}, J., {et~al.} 2009, \apj, 703, 2033

\bibitem[{{Reddy} {et~al.}(2010){Reddy}, {Erb}, {Pettini}, {Steidel}, \&
  {Shapley}}]{2010ApJ...712.1070R}
{Reddy}, N.~A., {Erb}, D.~K., {Pettini}, M., {Steidel}, C.~C., \& {Shapley},
  A.~E. 2010, \apj, 712, 1070

\bibitem[{{Reddy} \& {Steidel}(2009)}]{2009ApJ...692..778R}
{Reddy}, N.~A. \& {Steidel}, C.~C. 2009, \apj, 692, 778

\bibitem[{{Teplitz} {et~al.}(2013){Teplitz}, {Rafelski}, {Kurczynski},
  {et~al.}}]{2013AJ....146..159T}
{Teplitz}, H.~I., {Rafelski}, M., {Kurczynski}, P., {et~al.} 2013, \aj, 146,
  159

\bibitem[{{Wilkins} {et~al.}(2011){Wilkins}, {Bunker}, {Stanway}, {Lorenzoni},
  \& {Caruana}}]{2011MNRAS.417..717W}
{Wilkins}, S.~M., {Bunker}, A.~J., {Stanway}, E., {Lorenzoni}, S., \&
  {Caruana}, J. 2011, \mnras, 417, 717

\bibitem[{{Windhorst} {et~al.}(2011){Windhorst}, {Cohen}, {Hathi},
  {et~al.}}]{2011ApJS..193...27W}
{Windhorst}, R.~A., {Cohen}, S.~H., {Hathi}, N.~P., {et~al.} 2011, \apjs, 193,
  27

\end{thebibliography}

%
%

%
%
%


%
%
%


%
%
%


%
%
%


%
%
%
%
%
%
\begin{deluxetable}{lcccccccc} 
\tablecolumns{9} 
\tablewidth{0pc} 
\tablecaption{UV Continuum Slopes in UVUDF/HUDF12} 
\tablehead{ 
\colhead{ } & \colhead{ } & \multicolumn{4}{c}{UV Spectral Index, $\beta$} &\colhead{ } & \colhead{ }&  \colhead{ } \\ 
\cline{3-6} \\
\colhead{Category} & \colhead{$N$} & \colhead{$\left<\beta\right>$} &  \colhead{Biweight Mean} &\colhead{Median} & \colhead{$\sigma_\beta$} &\colhead{$d\beta/dM$} & \colhead{$d\beta/dz$}&  \colhead{$\rho$ (signif.)} \\ 
\colhead{(1)} & \colhead{(2)} & \colhead{(3)} & \colhead{(4)} & \colhead{(5)} & \colhead{(6)} & \colhead{(7)}  & \colhead{(8)} & \colhead{(9)}}
\startdata
\multicolumn{9}{c}{{\it Redshift Binned}}  \\
\hline \\
$1.0<z<2.0$ & 460 & $-1.382\pm0.002$ & $-1.80\pm0.04$ & $-1.83$ & 0.43 & $-0.09\pm0.05$ & $\cdots$ & $-0.16~(10^{-3})$\\
$2.0<z<3.0$ & 120 & $-1.791\pm0.005$ &  $-1.95\pm0.07$ &$-1.90$ & 0.36 & $-0.09\pm0.04$ & $\cdots$ & $-0.21~(0.021)$ \\
$3.0<z<4.0$ & 169 & $-1.624\pm0.003$ &  $-1.81\pm0.06$ &$-1.81$ & 0.37 & $-0.10\pm0.03$ & $\cdots$ & $-0.34~(10^{-5})$ \\
$4.0<z<7.7$ & 174 & $-1.668\pm0.006$ & $-1.90\pm0.05$ &$-1.91$ & 0.36 & $-0.23\pm0.04$ & $\cdots$ & $-0.37~(10^{-7})$ \\
\hline \\
\multicolumn{9}{c}{{\it Magnitude Binned}}  \\
\hline \\
$-15<M_{\textnormal{UV}}<-14$   &   23 & $-1.343\pm0.071$ &	$-1.72\pm0.31$	& $-1.72$ & 0.72  & $\cdots$ &$ 1.515\pm0.492$ &$ -0.28~(0.19)$\\
$-16<M_{\textnormal{UV}}<-15$   &   90 & $-1.695\pm0.023$ &	$-1.83\pm0.10$	& $-1.88$ & 0.45 & $\cdots$ & $-0.138\pm0.079$ &$ -0.23~(0.03)$ \\
$-17<M_{\textnormal{UV}}<-16$   &  152 & $-1.752\pm0.011$ &	$-1.90\pm0.06$	 & $-1.89$ & 0.37 & $\cdots$ & $-0.360\pm0.025$ & $-0.33~(10^{-5})$ \\
$-18<M_{\textnormal{UV}}<-17$   &  273 & $-1.796\pm0.005$ &	$-1.91\pm0.04$	 & $-1.91$ & 0.37 & $\cdots$ & $-0.052\pm0.006$ & $-0.28~(10^{-6})$\\
$-19<M_{\textnormal{UV}}<-18$   &  243 & $-1.518\pm0.003$ &	$-1.81\pm0.05$	& $-1.79$ & 0.41 & $\cdots$ & $-0.189\pm0.003$ & $-0.36~(10^{-8})$ \\
$-20<M_{\textnormal{UV}}<-19$   &  115 & $-1.464\pm0.003$  &	$-1.75\pm0.06$	& $-1.76$ & 0.35 & $\cdots$ & $-0.148\pm0.002$ & $-0.25~(0.006)$\\
\hline \\
\multicolumn{9}{c}{{\it Redshift \& Magnitude Binned}}  \\
\hline \\
\multicolumn{1}{l}{This Work}  & & & & & & & \\
$M_{\textnormal{UV}}\sim -17.5$ & $\cdots$ & $-1.92\pm0.04$ & $-1.96\pm0.16$ & $-1.93$ & 0.17 & $-0.13\pm0.02^a$ &  ${\bf-0.07\pm0.03}$ & $-0.71~(0.05)^c$\\
$M_{\textnormal{UV}}\sim -19.5$ & $\cdots$ & $-1.78\pm0.05$ & $-1.77\pm0.20$ & $-1.75$ & 0.12 & $-0.13\pm0.02^a$ &  ${\bf-0.04\pm0.03}$  & $-0.60~(0.28)^c$\\
\hline \\
\multicolumn{1}{l}{Combined Literature}  & & & & & & & \\
$M_{\textnormal{UV}}\sim -17.5$  & $\cdots$ & $-1.846\pm0.014$ & $-2.00\pm0.11$ & $-1.93$ & 0.25 & $-0.11\pm0.01^a$ &  ${\bf-0.09\pm0.01}^b$ & $-0.69~(10^{-4})^c$\\
$M_{\textnormal{UV}}\sim -19.5$ & $\cdots$ & $-1.522\pm0.002$ & $-1.92\pm0.09$ & $-1.90$ & 0.22 & $-0.11\pm0.01^a$ &  ${\bf-0.06\pm0.01}^b$  & $-0.78~(10^{-5})^c$\\
\vspace{0.010in}\\
\hline
\vspace{0.010in}\\
Entire Sample   & 923 & $-1.520\pm0.002$ & $-1.84\pm0.03$ & $-1.85$ & 0.41 &  $-0.082\pm0.002$  & $-0.112\pm0.001$ & $-0.16~(10^{-6})^c$ \\
\enddata

\tablecomments{Column (1) indicates the sample luminosity/redshift category.  Column (2) indicates the number of galaxies.  Column (3) indicates the inverse variance weighted average $\beta$ and standard error of the mean.  Column (4) is the Biweight Mean $\beta$. Column (5) indicates median $\beta$.  Column (6) is the the standard deviation of $\beta$.  Columns (7,8) are best fit slopes from weighted least squares.  Numbers in boldface have $\chi^2$ probabilities $ > 0.1$.  Column (9) is the Spearman correlation coefficient, $\rho$ (significance in parentheses).  $^a$Inverse variance weighted average of values shown in Figure \ref{FIGURE:dbdmVsMUVLiterature}.  standard deviation = 0.08.  $^b$$F$-test significance = $10^{-4}$ ($10^{-5}$) for $M_{\textnormal{UV}} \sim -17.5~(-19.5)$ combined literature data.  $^c$ Refers to $\beta(z)$.}
\label{TABLE:ResultsSummary}
\end{deluxetable} 

\end{document}